\documentclass[conference, letter]{IEEEtran}
\IEEEoverridecommandlockouts
\usepackage{cite}
\usepackage{amsmath,amssymb,amsfonts}
\usepackage{algorithmic}
\usepackage{graphicx}
\usepackage{textcomp}
\usepackage{xcolor}
\usepackage{comment}
\usepackage{mathtools}
\usepackage[cal=cm]{mathalfa}
\usepackage{cite}
\usepackage{siunitx}
\usepackage{wrapfig}
\usepackage{tikz}
\usepackage{pgfplots}
\usepackage{bbm}

\usetikzlibrary{external}
\ifCLASSOPTIONcompsoc
 \usepackage[caption=false,font=normalsize,labelfon
 t=sf,textfont=sf]{subfig}
 \else
 \usepackage[caption=false,font=footnotesize]{subfi
 g}
 \fi

\def\BibTeX{{\rm B\kern-.05em{\sc i\kern-.025em b}\kern-.08em
    T\kern-.1667em\lower.7ex\hbox{E}\kern-.125emX}}

\DeclareMathOperator{\arctanh}{tanh^{--1}}
\DeclareMathOperator{\sign}{sgn}

\usepackage{accents}
\newcommand{\dbtilde}[1]{\accentset{\approx}{#1}}

\definecolor{new_blue}{RGB}{93,169,233}
\definecolor{new_green}{HTML}{0FA3B1}
\definecolor{new_red}{HTML}{DD2D4A}
\definecolor{new_braun}{HTML}{79745C}
\definecolor{new_cherry}{HTML}{880D1E}
\definecolor{new_fildisi}{HTML}{F5EFED}
\definecolor{new_turqouise}{HTML}{508991}
\definecolor{new_purple}{HTML}{473198}
\definecolor{new_gray}{HTML}{67597A}

\definecolor{new_purple_soft}{HTML}{ACA2D2}
\definecolor{new_gray_soft}{HTML}{BCB5C4}
\definecolor{new_red_soft}{HTML}{DE828E}
\definecolor{KITred}{rgb}{.63,.13,.13}
\definecolor{KITblue}{rgb}{.27,.39,.66}
\definecolor{KITgreen}{rgb}{0,.59,.51}

\begin{document}

\title{Log-Log Domain Sum-Product Algorithm for Information Reconciliation in Continuous-Variable Quantum Key Distribution\\

\thanks{This work was funded by the German Federal Ministry of Education and Research (BMBF) under grant agreement 16KISQ056 (DE-QOR).}
}

\author{\IEEEauthorblockN{ Erdem Eray Cil and Laurent Schmalen}
\IEEEauthorblockA{{Communications Engineering Lab}, {Karlsruhe Institute of Technology},
76187 Karlsruhe, Germany \\
\texttt{erdem.cil@kit.edu}}
}

\maketitle

\begin{abstract}

In this paper, we present a novel log-log domain sum-product algorithm (SPA) for decoding low-density parity-check (LDPC) codes in continuous-variable quantum key distribution (CV-QKD) systems. This algorithm reduces the fractional bit width of decoder messages, leading to a smaller memory footprint and a lower resource consumption in hardware implementation. We also provide practical insights for fixed-point arithmetic and compare our algorithm with the conventional SPA in terms of performance and complexity. Our results show that our algorithm achieves comparable or better decoding accuracy than the conventional SPA while saving at least $25\%$ of the fractional bit width. 
\end{abstract}


\section{Introduction}

Quantum computing has made remarkable advances in recent years, challenging the security of conventional data encryption schemes that often rely on the hardness of factorizing large numbers \cite{Djordjevic_book}. A promising solution is quantum key distribution (QKD), which leverages the laws of physics to generate and distribute secret keys for symmetric encryption. Among the various QKD techniques, continuous-variable quantum key distribution (CV-QKD) has demonstrated great potential for long-distance secure communication, as shown, e.g., by the experimental achievement of a key exchange over more than 200 km \cite{208kmCV-QKD}.

However, the practical implementation of CV-QKD faces several challenges, particularly in the information reconciliation phase, where the parties involved aim to extract a common raw key from their measurements. This critical step typically makes use of multi-edge type (MET) low-density parity-check~(LDPC) codes \cite{9562244, Leverrier_CV_QKD_gaussian_mod, Mani_2021}, which require long codewords~\cite{9562244} and are sensitive to changes in decoding algorithms. Simplified belief propagation algorithms like the scaled min-sum algorithm often exhibit suboptimal performance \cite{Milicevic_2018, Cil24OFC}.

The requirements of an LDPC decoder in long-distance CV-QKD systems pose practical hardware challenges. To address these challenges, GPU-based decoders are commonly used due to their high parallelism and floating-point precision, which improve performance and throughput~\cite{Milicevic_2018, 208kmCV-QKD}. However, these decoders have drawbacks such as high energy consumption and host dependence, limiting their commercial viability. As an alternative, FPGA-based decoders offer high parallel processing capabilities with lower energy consumption and host independence. However, CV-QKD systems require high precision in the fixed-point representation of decoder messages, which increases the memory footprint of the decoder \cite{yangFPGABasedLDPCDecoder2021}. To address this issue, a two-stage decoder architecture is proposed in \cite{zhouHighthroughputDecoderQuasicyclic2022}. By decreasing the fractional bit width, this architecture can reduce the memory footprint. However, it requires additional decoding iterations, leading to lower throughput.

In this paper, we propose a novel log-log domain sum-product algorithm (SPA) that reduces the precision of message representations by at least $25\%$ without increasing the decoding iterations or decoding complexity compared to the conventional SPA. Section~\ref{sec:background} provides the background on forward error correction and information reconciliation for CV-QKD systems. In Section~\ref{sec:app_CN_update}, we derive an approximation for the check node (CN) update of the SPA, streamlining CN processing. Section~\ref{sec:log_log_SPA} introduces the log-log domain SPA, and in Section \ref{sec:fixed_point_representation}, we briefly explain its implementation in fixed-point arithmetic. Section~\ref{sec:results} offers a comparative analysis between SPA and log-log domain SPA, demonstrating the reduction in memory footprint achieved by the latter.

\section{Background}
\label{sec:background}
\subsection{Forward Error Correction in CV-QKD}

QKD allows two parties, Alice and Bob, to establish a shared secret by exchanging and measuring quantum states, while facing an adversary, Eve, who can access and manipulate the quantum channel. To generate a common bit string, which serves as raw key material, one of the parties, either Alice or Bob, uses forward error correction techniques to correct their string according to the other party’s string. This information reconciliation process is called reverse reconciliation when Alice corrects her string, and direct reconciliation otherwise. For CV-QKD, reverse reconciliation enables long-distance operation~\cite{silberhornContinuousVariableQuantum}.

We assume that the quantum channel carrying quantum states is an additive white Gaussian noise (AWGN) channel. The measured quantum state $r$ is the sum of the transmitted quantum state $t$ and the noise $n$, which has a variance of $\sigma_n^2$. Thus, the channel model is $ r = t + n $.

In reverse reconciliation, Bob measures the quantum states and generates the raw key material. He then sends this material to Alice through a noiseless classical public channel. To reduce the amount of information that an eavesdropper, Eve, can obtain from the message, Bob maps the random binary key material, denoted as $C$, to BPSK symbols and modulates them onto the measured quantum states before transmission. The message he sends through the noiseless public channel is $m = (-1)^C r$.

Then, Alice receives the message and divides it by $t$ to recover $C$ from $m$:
\begin{align}
    y &= \frac{m}{t} = \frac{(-1)^C (t + n)}{t} \nonumber\\
    &= (-1)^C + \frac{n}{t} \nonumber \\
    &= (-1)^C + n' . \label{eqn:channel_model}   
\end{align}
Assuming BPSK modulation for the transmitted quantum states, $n'$ follows a Gaussian distribution with zero mean and variance $\sigma_n^2$. Therefore, the synthetic channel after the division is a binary-input AWGN (BI-AWGN) channel.

If the transmitted quantum states are not BPSK modulated, we can still assume the synthetic channel to be BI-AWGN channel, if we perform multidimensional reconciliation~\cite{leverrierMultidimensionalReconciliationContinuousvariable2008}.

The transmit optical powers are kept low to remain in the quantum regime, and the received signal-to-noise ratio (SNR) values are small due to the long transmission distance in long-distance CV-QKD systems. Thus, we need to apply low-rate codes optimized for BI-AWGN channels to recover the key material $C$ in (\ref{eqn:channel_model}).

The secret key rate (SKR) is an important metric for CV-QKD because it quantifies the system's secure key generation rate. For reverse reconciliation, the SKR reads 
\begin{align}
\mathrm{SKR} &= (1 - \mathrm{FER})(\beta I_{AB} - \chi_{BE}) \label{eqn:SKR} \\
&= (1 - \mathrm{FER})(R - \chi_{BE}). \nonumber
\end{align}
Here, $I_{AB} \geq 0$ denotes the capacity of the quantum channel, $\chi_{BE} \geq 0 $ represents the Holevo bound on the information leaked to Eve, $\beta $ is the reconciliation efficiency defined as the ratio of the code rate $R$ to $I_{AB}$, and the FER is the frame error rate after the decoding. According to (\ref{eqn:SKR}), $\beta$ must be larger than \mbox{$ \chi_{BE} / I_{AB} $} to achieve a non-zero SKR. Thus, the system operates close to the channel capacity, and high FER values, on the order of $0.1$, are deemed acceptable \cite{Milicevic_2018}.

\subsection{Multi-Edge Type LDPC Codes for CV-QKD}

MET-LDPC codes with a cascade structure are suitable for long-distance CV-QKD systems that operate in low SNR quantum channels, as they can achieve near-capacity performance~\cite{208kmCV-QKD, Richardson:95821}. These codes have a graph structure that consists of two sub-graphs: one sub-graph represents a high-rate code, and the other sub-graph contains degree-1 variable nodes (VN) similar to low-density generator matrix codes. The connection between the sub-graphs is established by specific, optimized edges.

\begin{figure}
    \subfloat[]{
    \resizebox{0.22\textwidth}{!}{    
    \tikzsetnextfilename{output-figure0}%
    \begin{tikzpicture}
\draw[fill = black] (0,-0.75) circle (0.3cm);

\draw[thick]  (2,0) circle (0.3cm);
\draw[thick]  (1.7,0) -- (2.3,0);
\draw[thick]  (2,0.3) -- (2,-0.3);
\draw[thick] (1,0) -- (1.7,0);
\draw[thick]  (2.3,0) -- (3,0);
\node at (2.65,0.35) {\LARGE 3};
\node at (1.35,0.35) {\LARGE 2};

\draw[thick,dashed,KITred] (1,3.2) rectangle (3,0.8);
\draw[thick]  (2,1.5) circle (0.3cm);
\draw[thick]  (1.7,1.5) -- (2.3,1.5);
\draw[thick]  (2,1.8) -- (2,1.2);
\draw[thick] (1,1.5) -- (1.7,1.5);
\draw[thick]  (2.3,1.5) -- (3,1.5);
\node at (2.7,1.1) {\Large \textcolor{KITred}{14}};
\draw[fill = black] (2,2.5) circle (0.3cm);
\draw[thick]  (2,1.8) -- (2,2.2);
\node at (2.65,1.85) {\LARGE 3};

\draw[thick,dashed,KITred] (1,-0.8) rectangle (3,-3.2);
\draw[thick]  (2,-1.5) circle (0.3cm);
\draw[thick]  (1.7,-1.5) -- (2.3,-1.5);
\draw[thick]  (2,-1.2) -- (2,-1.8);
\draw[thick] (1,-1.5) -- (1.7,-1.5);
\draw[thick]  (2.3,-1.5) -- (3,-1.5);
\node at (2.7,-2.9) {\Large \textcolor{KITred}{6}};
\draw[fill = black] (2,-2.5) circle (0.3cm);
\draw[thick]  (2,-1.8) -- (2,-2.2);
\node at (2.65,-1.15) {\LARGE 2};

\draw[thick,dashed,KITred] (1,-3.3) rectangle (3,-5.7);
\draw[thick]  (2,-4) circle (0.3cm);
\draw[thick]  (1.7,-4) -- (2.3,-4);
\draw[thick]  (2,-3.7) -- (2,-4.3);
\draw[thick] (1,-4) -- (1.7,-4);
\draw[thick]  (2.3,-4) -- (3,-4);
\node at (2.7,-5.4) {\Large \textcolor{KITred}{6}};
\draw[fill = black] (2,-5) circle (0.3cm);
\draw[thick]  (2,-4.3) -- (2,-4.7);
\node at (2.65,-3.65) {\LARGE 4};

\draw[fill = black] (4,-0.75) circle (0.3cm);

\draw[thick,dashed,KITred] (5,1.95) rectangle (8.2,0.55);
\draw[thick]  (6,1.25) circle (0.3cm);
\draw[thick]  (5.7,1.25) -- (6.3,1.25);
\draw[thick]  (6,1.55) -- (6,0.95);
\draw[fill = black] (7.4,1.25) circle (0.3cm);
\draw[thick]  (5,1.25) -- (5.7,1.25);
\draw[thick]  (6.3,1.25) -- (7.1,1.25);
\node at (7.9,0.85) {\Large \textcolor{KITred}{18}};
\node at (5.35,1.6) {\LARGE 3};

\draw[thick,dashed,KITred] (5,-1.45) rectangle (8.2,-0.05);
\draw[thick]  (6,-0.75) circle (0.3cm);
\draw[thick]  (5.7,-0.75) -- (6.3,-0.75);
\draw[thick]  (6,-1.05) -- (6,-0.45);
\draw[fill = black] (7.4,-0.75) circle (0.3cm);
\draw[thick]  (5,-0.75) -- (5.7,-0.75);
\draw[thick]  (6.3,-0.75) -- (7.1,-0.75);
\node at (7.9,-1.15) {\Large \textcolor{KITred}{35}};
\node at (5.35,-0.4) {\LARGE 2};

\draw[thick,dashed,KITred] (5,-3.45) rectangle (8.2,-2.05);
\draw[thick]  (6,-2.75) circle (0.3cm);
\draw[thick]  (5.7,-2.75) -- (6.3,-2.75);
\draw[thick]  (6,-3.05) -- (6,-2.45);
\draw[fill = black] (7.4,-2.75) circle (0.3cm);
\draw[thick]  (5,-2.75) -- (5.7,-2.75);
\draw[thick]  (6.3,-2.75) -- (7.1,-2.75);
\node at (7.9,-3.15) {\Large \textcolor{KITred}{19}};
\node at (5.35,-2.4) {\LARGE 4};

\draw[thick] (0,-0.75) -- (1,0);
\draw[thick] (0,-0.75) -- (1,1.5);
\draw[thick] (0,-0.75) -- (1,-1.5);
\draw[thick] (0,-0.75) -- (1,-4);

\draw[thick] (4,-0.75) -- (3,0);
\draw[thick] (4,-0.75) -- (3,1.5);
\draw[thick] (4,-0.75) -- (3,-1.5);
\draw[thick] (4,-0.75) -- (3,-4);

\draw[thick] (4,-0.75) -- (5,1.25);
\draw[thick] (4,-0.75) -- (5,-0.75);
\draw[thick] (4,-0.75) -- (5,-2.75);
\end{tikzpicture}}
    \label{fig:protograph_R_0p01}
    }
    \subfloat[]{
    \raisebox{1cm}{\resizebox{0.22\textwidth}{!}{
    \tikzsetnextfilename{output-figure1}
    \begin{tikzpicture}
\draw[fill = black] (0,0) circle (0.3cm);

\draw[thick] (2,0) circle (0.3cm);
\draw[thick]  (1.7,0) -- (2.3,0);
\draw[thick]  (2,0.3) -- (2,-0.3);
\draw[thick] (1,0) -- (1.7,0);
\draw[thick]  (2.3,0) -- (3,0);
\node at (2.65,0.35) {\LARGE 3};
\node at (1.35,0.35) {\LARGE 2};

\draw[thick]  (2,1.5) circle (0.3cm);
\draw[thick]  (1.7,1.5) -- (2.3,1.5);
\draw[thick]  (2,1.8) -- (2,1.2);
\draw[thick] (1,1.5) -- (1.7,1.5);
\draw[thick]  (2.3,1.5) -- (3,1.5);
\draw[fill = black] (2,2.5) circle (0.3cm);
\draw[thick]  (2,1.8) -- (2,2.2);
\node at (2.65,1.85) {\LARGE 3};

\draw[thick]  (2,-1.5) circle (0.3cm);
\draw[thick]  (1.7,-1.5) -- (2.3,-1.5);
\draw[thick]  (2,-1.2) -- (2,-1.8);
\draw[thick] (1,-1.5) -- (1.7,-1.5);
\draw [thick] (2.3,-1.5) -- (3,-1.5);
\draw[fill = black] (2,-2.5) circle (0.3cm);
\draw[thick]  (2,-1.8) -- (2,-2.2);
\node at (2.65,-1.15) {\LARGE 2};

\draw[fill = black] (4,0) circle (0.3cm);

\draw[thick,dashed, KITred] (5,1.7) rectangle (8.2,0.3);
\draw[thick]  (6,1) circle (0.3cm);
\draw[thick]  (5.7,1) -- (6.3,1);
\draw[thick]  (6,1.3) -- (6,0.7);
\draw[fill = black] (7.4,1) circle (0.3cm);
\draw[thick]  (5,1) -- (5.7,1);
\draw[thick]  (6.3,1) -- (7.1,1);
\node at (7.9,0.6) {\Large \textcolor{KITred}{5}};
\node at (5.35,1.35) {\LARGE 3};

\draw[thick]  (6,-1) circle (0.3cm);
\draw[thick]  (5.7,-1) -- (6.3,-1);
\draw[thick]  (6,-1.3) -- (6,-0.7);
\draw[fill = black] (7.4,-1) circle (0.3cm);
\draw[thick]  (5,-1) -- (5.7,-1);
\draw[thick]  (6.3,-1) -- (7.1,-1);
\node at (5.35,-0.65) {\LARGE 2};

\draw[thick] (0,0) -- (1,0);
\draw[thick] (0,0) -- (1,1.5);
\draw[thick] (0,0) -- (1,-1.5);

\draw[thick] (4,0) -- (3,0);
\draw[thick] (4,0) -- (3,1.5);
\draw[thick] (4,0) -- (3,-1.5);

\draw[thick] (4,0) -- (5,1);
\draw[thick] (4,0) -- (5,-1);

\end{tikzpicture}
    }
    }
    \label{fig:protograph_R_0p1}
    }
\caption{Protographs of the optimized TBP-LDPC codes \cite{9562244} of rates (a)~\mbox{$R=0.01$} and (b) $R=0.1$. The (red) numbers at the lower right in the boxes represents the optimized repetition values for the subgraph in the box. }
\label{fig:protographs}
\end{figure}
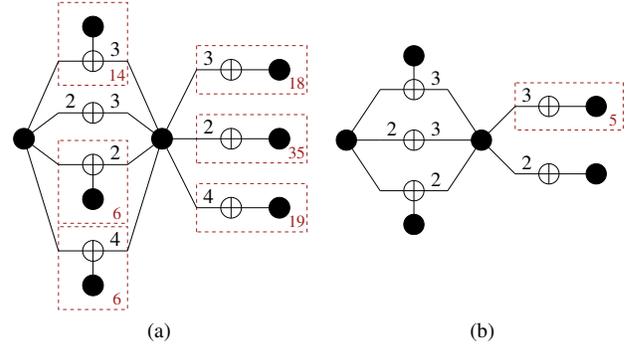

To simplify the design of low-rate MET-LDPC codes, the concept of type-based protograph LDPC (TBP-LDPC) codes is introduced in \cite{9562244}. The protographs for TBP-LDPC codes with rates $R=0.01$ and $R=0.1$ are optimized and presented in \figurename~\ref{fig:protographs}, following the notation from \cite{9562244}. The parity check matrices of the codes used in the simulations are obtained by lifting these protographs.

In this paper, the following notation is employed: \mbox{$E_s/N_0=1/(2\sigma_n^2)$} represents the SNR of the BI-AWGN channel. Log-likelihood ratio (LLR) values are denoted by $L$, where \mbox{$L=\alpha\gamma$}, with $\alpha$ representing the sign and $\gamma$ the magnitude of the LLR. The logarithm of the magnitude of LLR values is denoted by $\Tilde{L}$, hence $\gamma=\exp{(\Tilde{L})}$. $L_{\mathrm{ch},i}$ stands for the channel LLR value for the VN $i$. For an LDPC code with parity check matrix $\mathbf{H}$, the set $\mathcal{N}(i)=\{ j: \mathbf{H}_{j,i} =1 \}$ denotes the connection set of the VN $i$, whereas the set $\mathcal{M}(j)=\{ i: \mathbf{H}_{j,i} =1 \}$ denotes the connection set of the CN~$j$. $L^{[\mathrm{c}]}_{i \gets j }$ and $L^{[\mathrm{v}]}_{j \to i }$ represent the messages from CN/VN $j$ to VN/CN $i$, respectively. $N \in \mathbb{N}$ denotes the blocklength of the code.   

\subsection{Sum Product Algorithm}
\label{sec:SPA}

The SPA stands out as the most effective practical algorithm for decoding LDPC codes. To ensure numerical stability, LLR values are used as the messages in the SPA decoding. The node update rules for the SPA can be expressed as:
 
\begin{enumerate}
    \item Initialization: for each VN $i$ with $i \in \{ 1 \ldots N \} $ and $j \in \mathcal{N}(i)$, set $L^{[\mathrm{v}]}_{i \to j } = L_{\mathrm{ch},i}$. 
    \item CN update: for each CN $j$ and \mbox{$i \in \mathcal{M}(j)$}, calculate
\begin{align}
   & L^{[\mathrm{c}]}_{i \gets j } = 2 \hspace{-0.5cm}\prod_{k\in \mathcal{M}(j)\backslash \{i \}} \hspace{-0.4cm} \alpha_k \cdot \arctanh \left( \hspace{-0.05cm}\prod_{\ell\in \mathcal{M}(j)\backslash \{i\}} \hspace{-0.4cm} \tanh\left(\frac{\gamma_\ell}{2}\right) \right) \label{eqn:full_SPA}
\end{align}
where $L^{[\mathrm{v}]}_{\ell \to j } = \alpha_\ell \gamma_\ell$.
    \item VN update: for each VN $i$  and \mbox{$j \in \mathcal{N}(i)$}, calculate
\begin{align}
   & L^{[\mathrm{v}]}_{i \to j } = L_{\mathrm{ch},i} +\sum_{k\in \mathcal{N}(i)\backslash \{j \}} L^{[\mathrm{c}]}_{i \gets k } \nonumber 
\end{align}
    \item A posteriori LLR: for each VN $i$ with $i \in \{ 1 \ldots N \} $, calculate
\begin{align}
    L^{[\mathrm{total}]}_{i } &= L_{\mathrm{ch},i} +\sum_{k\in \mathcal{N}(i)} L^{[\mathrm{c}]}_{i \gets k } \nonumber \\
    \hat{x}_i &= \frac{1- \sign{(L^{[\mathrm{total}]}_{i })}}{2} \nonumber
\end{align}
    \item Stopping Criteria:  if $\bold{H\hat{x}}^T=\bold{0}$, with $\bold{\hat{x}} = (\hat{x}_1, \ldots, \hat{x}_N )$ or the number of decoding iterations equals the limit, stop. Otherwise, go to Step 2.
\end{enumerate}

\section{The Approximate CN Update Equation}
\label{sec:app_CN_update}

In this section, we revisit the approximation for the SPA CN update equation from \cite{maganaDifferentPerspectiveApproach2012}, originally designed for high rate codes, demonstrating its applicability for low rate codes. To achieve this, we start with the CN update equation for a degree-3 CN and then generalize it to higher degrees.

The box-plus operator for input LLRs $L_1=\alpha_1\gamma_1$ and \mbox{$L_2=\alpha_2\gamma_2$}, where $\alpha_i$ and $\gamma_i$ represent the sign and the magnitude of the LLR respectively, can be expressed as~\mbox{\cite[Ch. 5]{Ryan2009}}:
\begin{align}
    L_1 \boxplus L_2 &= \nonumber \\ \alpha_1\alpha_2 & \Bigg( \min(\gamma_1,\gamma_2) + \underbrace{\log{\left(\frac{1+\mathrm{e}^{-\big\lvert\gamma_1+\gamma_2\big\lvert}}{1+\mathrm{e}^{-\big\lvert\gamma_1-\gamma_2\big\lvert}}\right)}}_{\eqqcolon \ \mathrm{s}(\gamma_1,\gamma_2)} \Bigg) . \label{eqn:boxplus_section}
\end{align}
Without loss of generality, we assume that \mbox{$\gamma_\mathrm{min}  \coloneqq  \gamma_1  \leq   \gamma_\mathrm{max} \coloneqq \gamma_2$ and $\gamma_\mathrm{max} = {d}\gamma_\mathrm{min}$} for any  \mbox{$1 \leq {d} \in \mathbb{R}$}. Under this assumption, the correction term $\mathrm{s}(\gamma_1,\gamma_2) = \mathrm{s}(\gamma_\mathrm{min},\gamma_\mathrm{max})$ can be reformulated as follows: 
\begin{align}
     \mathrm{s} (\gamma_\mathrm{min}, \gamma_\mathrm{max}) & = \log{(1\hspace{-0.1cm}+\hspace{-0.05cm}\mathrm{e}^{-({d}+1)\gamma_\mathrm{min}})}\hspace{-0.1cm} - \hspace{-0.1cm}\log{(1\hspace{-0.1cm}+\hspace{-0.05cm}\mathrm{e}^{-({d}-1)\gamma_\mathrm{min}})} \nonumber\\
    &\overset{(i)}{=}\sum_{k=1}^{\infty} \frac{(-1)^{k-1}}{k} \mathrm{e}^{-k{d}\gamma_\mathrm{min}} \nonumber  \left(\mathrm{e}^{-k\gamma_\mathrm{min}}-\mathrm{e}^{k\gamma_\mathrm{min}}\right) \nonumber\\
    &\overset{(ii)}{\approx} 2 \sum_{k=1}^{\infty} (-1)^{k}\mathrm{e}^{-k{d}\gamma_\mathrm{min}}\gamma_\mathrm{min} \nonumber \\ 
    &\overset{(iii)}{=} -\frac{2\ \gamma_\mathrm{min}}{1+\mathrm{e}^{{d}\gamma_\mathrm{min}}} \nonumber \\
    & \ = -\frac{2\ \gamma_\mathrm{min}}{1+\mathrm{e}^{\gamma_\mathrm{max}}} \cdot\label{eqn:correction_term}
\end{align}
Here, step $(i)$ involves expressing the logarithmic terms using the Taylor series. In step $(ii)$, by assuming small values of $k\gamma_\mathrm{min}$, we employ $\exp(x)\approx 1+x$. This assumption is reasonable since the density of the degree-1 VNs in typical low-rate codes used in CV-QKD is high. To give an example, in the TBP-LDPC code of rate 0.01 in \figurename~\ref{fig:protograph_R_0p01}, 98.9\% of the CNs are connected to degree-1 VNs. It is important to note that this approximation holds for small values of $k\gamma_\mathrm{min}$, and the exponential decay $\exp{(-k{d}\gamma_\mathrm{min})}$ aids in reducing the error between the actual function and its approximation for large values of $k\gamma_\mathrm{min}$. In step $(iii)$, we employ the  geometric series $1/(1+x)=\sum_{k=0}^\infty (-x)^k$.

Replacing the correction term in (\ref{eqn:boxplus_section}) with the approximate correction term (\ref{eqn:correction_term}) yields 
\begin{align}
    L_1 \boxplus L_2 &\approx \alpha_1\alpha_2 \left( \gamma_\mathrm{min} - \frac{2\ \gamma_\mathrm{min}}{1+\mathrm{e}^{\gamma_\mathrm{max}}} \right) \nonumber \\
    &= \alpha_1\alpha_2 \gamma_\mathrm{min} \tanh\left(\frac{\gamma_\mathrm{max}}{2}\right) \cdot  \nonumber %
\end{align}
Hence, the output LLR of the CN $j$ to the VN $i$ for  $i \in \mathcal{M}(j)$  can be generalized as:
\begin{equation}
    L^{[\mathrm{c}]}_{i \gets j } \approx  \prod_{k\in \mathcal{M}(j)\backslash \{i \}}\hspace{-0.5cm} \alpha_k \cdot \gamma_m \cdot \left(\prod_{\ell\in \mathcal{M}(j)\backslash \{i ,m\}} \hspace{-0.5cm} \tanh\left(\frac{\gamma_\ell}{2}\right) \right) \label{eqn:app_CN_update},
\end{equation}
where $m$ is defined to be:
\begin{align}
    m = \mathrm{arg} \ \mathrm{min}_{k \in \mathcal{M}(j)´\backslash \{i\} } \ |L^{[\mathrm{v}]}_{k \to j }| . \label{eqn:defn_m} 
\end{align}

The approximation (\ref{eqn:app_CN_update}) shows how each incoming LLR value affects the output value. Specifically, the $\tanh$ function saturates for high-magnitude LLR values, so they can be ignored in the computation. On the other hand, low-magnitude values scale the output value by $\tanh{(\gamma_\ell/2)}$, so they have a proportional impact. This feature can be used to improve the performance of scaled min-sum decoding by choosing more accurate scaling coefficients, as explained in \cite{maganaDifferentPerspectiveApproach2012, Cil24OFC}.

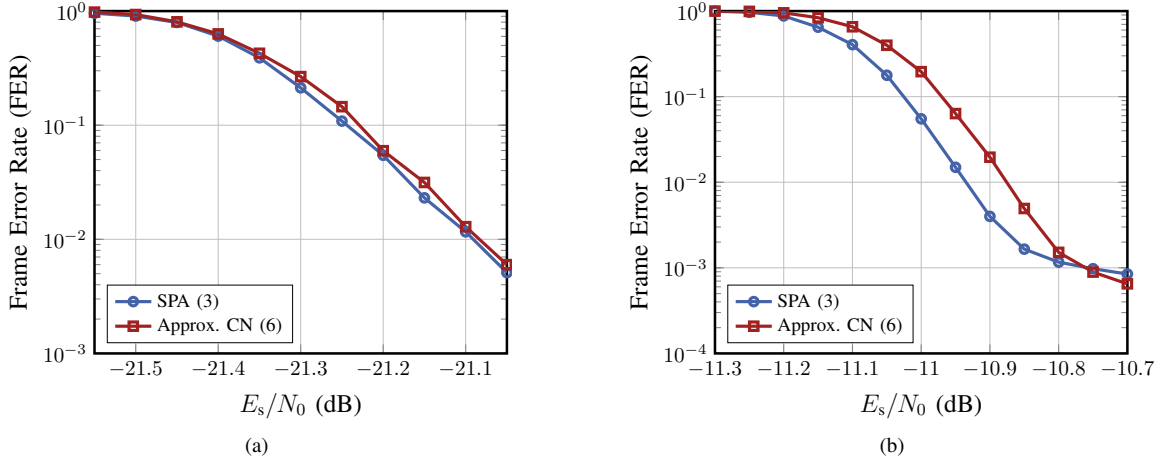
\begin{figure*}[t!]
\centering
\subfloat[ ]{
\tikzsetnextfilename{output-figure2}%
\begin{tikzpicture}[every pin/.style={fill=white},scale=0.8]
  \begin{axis}[
    grid=major,
    xlabel={ \large \text{$E_\text{s}/N_0$} (dB)},
    ylabel={\large Frame Error Rate (FER)},
    ymode=log,
    legend pos=south west,
    grid=major,
    ymax=1,
    ymin=0.001,
    legend style={align=left},
    legend cell align=left, 
    xmin=-21.55,
    xmax=-21.05,
    axis line style =very thick,
  ]
\addplot[KITblue,mark=o, line width=1.5] table [x=SNR,y=FER] {data/R_0p01_SPA_double_precision.txt};
\addlegendentry{ \small SPA (\ref{eqn:full_SPA})} %
\addplot[KITred,mark=square, line width=1.5] table [x=SNR,y=FER] {data/R_0p01_CN_approximation_floating_point.txt};
\addlegendentry{ \small Approx. CN (\ref{eqn:app_CN_update})} %
\end{axis}
\end{tikzpicture}
\label{fig:comp_R_0p01}
}
\hfil
\subfloat[ ]{
\tikzsetnextfilename{output-figure3}%
\begin{tikzpicture}[every pin/.style={fill=white},scale=0.8]%
\begin{axis}[
    xlabel={\large \text{$E_\text{s}/N_0$} (dB)},
    ylabel={\large{Frame Error Rate (FER)}},
    ymode=log,
    legend pos=south west,
    grid=major,
    ymax=1,
    ymin=0.0001,
    legend style={align=left},
    xmin=-11.3,
    xmax=-10.7,
    legend cell align=left, 
    axis line style =very thick,
]
\addplot[KITblue,mark=o, line width=1.5] table [x=SNR,y=FER] {data/R_0p1_SPA_double_precision.txt};
\addlegendentry{ \small SPA (\ref{eqn:full_SPA})} %
\addplot[KITred,mark=square, line width=1.5] table [x=SNR,y=FER] {data/R_0p1_CN_approximation_floating_point.txt};
\addlegendentry{ \small Approx. CN (\ref{eqn:app_CN_update})} %
\end{axis}
\end{tikzpicture} 
\label{fig:comp_R_0p1}
}
\captionsetup{font=small}
\caption{Performance comparison of the SPA CN and the approximate CN update equations for the TBP-LDPC codes of rates (a) $R=0.01$ with $N=998400$ and (b) $R=0.1$ with $N=128000$}
\label{fig:app_comparison}
\vspace{-13pt}
\end{figure*}

In \figurename~\ref{fig:app_comparison}, we compare the performances of decoders utilizing the original CN update (\ref{eqn:full_SPA}) and the approximation (\ref{eqn:app_CN_update}). To assess the accuracy of the approximation across different SNR regimes, we employ two TBP-LDPC codes of rates $0.1$ and $0.01$ with the protographs given in \figurename~\ref{fig:protographs}.

In the ultra-low rate regime ($R=0.01$), as depicted in \figurename~\ref{fig:comp_R_0p01}, the approximation closely aligns with the original CN update equation, resulting in minimal performance degradation. However, in the low-rate scenario ($R=0.1$), the degradation is approximately \SI{0.05}{dB} at an FER of $0.1$, as can be seen in \figurename~\ref{fig:comp_R_0p1}. It is noteworthy that the performance of the approximate decoding surpasses that of the SPA in the error floor region for $R=0.1$, a phenomenon also observed in \cite{maganaDifferentPerspectiveApproach2012} for high rate codes.

\section{Log-log Domain SPA}
\label{sec:log_log_SPA}

Low-rate decoders need high precision for message representation in the decoder\cite{yangFPGABasedLDPCDecoder2021}. This can be seen from the approximate CN update equation. By using (\ref{eqn:app_CN_update}) and linearizing $\tanh$, the CN output LLR is proportional to the product of the incoming messages. The CN output LLR is an extrinsic message that helps to increase the bit reliability, but it is very small when the incoming messages are small in average. If the fixed-point representation is not precise enough, this extrinsic message is rounded to zero and the VN information will not improve. Therefore, low-magnitude messages require high precision, while high-magnitude messages can tolerate lower precision.

To implement decoders for ultra-low rate codes with rates around \mbox{$R=0.01$} on resource-limited FPGAs, the high precision requirement poses a significant challenge. Together with the high block lengths of the codes, it causes high memory consumption. Consequently, new decoding algorithms that can work with lower precision are needed. To this end, we propose a decoding algorithm that uses the logarithmic LLRs~\mbox{(log-LLR)}. The uniformly quantized log-LLR has the advantage of having high precision for the small LLR magnitudes and low precision for large ones. We refer to this algorithm as log-log domain SPA.

Utilizing (\ref{eqn:app_CN_update}), the log-log domain SPA can be articulated as follows:

\begin{enumerate}
    \item Initialization: for each VN $i$ with $i \in \{ 1 \ldots N \} $ and $j \in \mathcal{N}(i)$, set $\tilde{L}^{[\mathrm{v}]}_{i \to j } = \tilde L_{\mathrm{ch},i} = \ln(|L_{\mathrm{ch},i}|)$ and \mbox{${\alpha}^{[\mathrm{v}]}_{i \to j } =\alpha_i = \sign(L_{\mathrm{ch},i})$}. 
    \item CN update: for each CN $j$ and \mbox{$i \in \mathcal{M}(j)$}, calculate
\begin{align}
    \tilde{L}^{[\mathrm{c}]}_{i \gets j } &= \tilde{L}^{[\mathrm{v}]}_{m \to j } + \hspace{-0.8cm} \sum_{\ell\in \mathcal{M}(j)\backslash \{i, m\}} \hspace{-0.5cm} \ln \left(  \tanh{\left(\frac{\exp{(\tilde{L}^{[\mathrm{v}]}_{\ell \to j })}}{2}\right)} \right) \label{eqn:log_SPA}\\ 
    {\alpha}^{[\mathrm{c}]}_{i \gets j } &=  \prod_{k\in \mathcal{M}(j)\backslash \{i \}} {\alpha}^{[\mathrm{v}]}_{k \to j } , \nonumber
\end{align}
where $m$ is defined in (\ref{eqn:defn_m}).
    \item VN update: for each VN $i$  and \mbox{$j \in \mathcal{N}(i)$}, calculate
\begin{align}
    {L}^{[\mathrm{v}]}_{i \to j } &=  \alpha_i \exp(\tilde L_{\mathrm{ch},i}) +\hspace{-0.4cm} \sum_{k\in \mathcal{N}(i)\backslash \{j \}} \hspace{-0.3cm} {\alpha}^{[\mathrm{c}]}_{i \gets k } \exp(\tilde{L}^{[\mathrm{c}]}_{i \gets k }) \nonumber \\
   \tilde{L}^{[\mathrm{v}]}_{i \to j } &= \ln\left(\lvert {L}^{[\mathrm{v}]}_{i \to j } \lvert\right) \label{eqn:log_SPA_VN} \\
   {\alpha}^{[\mathrm{v}]}_{i \to j }  &= \sign\left( {L}^{[\mathrm{v}]}_{i \to j }  \right) \nonumber
\end{align}
    \item A posteriori LLR: for each VN $i$ with $i \in \{ 1 \ldots N \} $, calculate
\begin{align}
    L^{[\mathrm{total}]}_{i } &= \alpha_i\exp(\tilde L_{\mathrm{ch},i}) +\sum_{k\in \mathcal{N}(i)} {\alpha}^{[\mathrm{c}]}_{i \gets k }  \exp(\tilde{L}^{[\mathrm{c}]}_{i \gets k }) \nonumber \\
    \hat{x}_i &= \frac{1- \sign{(L^{[\mathrm{total}]}_{i })}}{2} \nonumber
\end{align}
    \item Stopping Criteria:  if $\bold{H\hat{x}}^T=\bold{0}$, with $\bold{\hat{x}} = (\hat{x}_1, \ldots, \hat{x}_N )$ or the number of decoding iterations equals the limit, stop. Otherwise, go to Step 2.
\end{enumerate}

During initialization, the logarithm of each channel LLR is computed. Subsequently, the log-LLR values and the sign of the channel LLRs are stored in the memory. The CN update~(\ref{eqn:log_SPA}) can be effectively approximated using a piecewise linear approach, eliminating the need for lookup tables. With this approach, (\ref{eqn:log_SPA}) becomes
\begin{align}
    \tilde{L}^{[\mathrm{c}]}_{i \gets j } &= \tilde{L}^{[\mathrm{v}]}_{m \to j } + \hspace{-0.5cm} \sum_{\ell\in \mathcal{M}(j)\backslash \{i, m\}} \hspace{-0.4cm} g{(\tilde{L}^{[\mathrm{v}]}_{\ell \to j })}, \label{eqn:CN_approx_piecewise} 
\end{align}
where $g(x)$ is defined as:
\begin{align}
    g(x) = \begin{cases}
        x - 0.694 & x\leq -0.76 \\
        0.833x-0.822 & -0.76 < x \leq 0.538 \\
        0.389x-0.583 & 0.538 < x \leq 1.414 \\
        0 & 1.414 < x .
    \end{cases} \label{eqn:piecewise_approx}
\end{align}
Since the scaling coefficients and biases in (\ref{eqn:piecewise_approx}) are constant, $g(x)$ can be easily implemented in hardware.

To compute the log-LLR value $\tilde{L}^{[\mathrm{v}]}_{i \to j}$ without using the $\exp$ function or calculating ${L}^{[\mathrm{v}]}_{i \to j}$, we use the log-sum/difference-exp functions $f_\pm(x,y)=\max(x,y) + \ln(1 \pm \exp(-|x-y|))$ in the VN update. In the pairwise processing of log-LLR values in (\ref{eqn:log_SPA_VN}), we need to evaluate either $ \ln(1 + \exp(-|x-y|)) $ or $ \ln(1 - \exp(-|x-y|)) $ depending on the sign bits of $x$ and $y$. We can store these functions as lookup tables for the decoding. The sign bit corresponding to this log-LLR value can be obtained in hardware by
\begin{align}
    \mathbbm{1}_{\{\alpha_x=\alpha_y\}}\alpha_x + \mathbbm{1}_{\{\alpha_x \neq \alpha_y\}} \left( \mathbbm{1}_{\{x \geq y\}} \alpha_x + \mathbbm{1}_{\{x < y\}} \alpha_y \right) , \label{eqn:sign_bit} 
\end{align}
where $\alpha_x$ and $\alpha_y$ are the sign bits corresponding to log-LLR values $x$ and $y$, and $\mathbbm{1}_{\{a=b\}}$ is the indicator function that outputs $1$ if the condition $a=b$ is satisfied, otherwise it evaluates to $0$. The implementation of (\ref{eqn:sign_bit}) in hardware can be achieved by using comparators and simple logic gates.

\newpage
\section{Fixed-Point Representation of Log-LLR Messages}
\label{sec:fixed_point_representation}

Fixed-point (FP) representation is preferred for numbers in FPGA, where each number consists of ($1$, $x$, $y$) bits. The first bit indicates the sign, the next $x$ bits represent the integer part, and the last $y$ bits denote the fractional part of the number. This allows us to represent values within the range $(-2^x, 2^x)$.

As discussed in the previous section, log-LLR messages are more robust to quantization than LLR messages, and thus need less precision in the FP representation. However, the sign of LLRs has to be stored along with the log-LLR values, which adds one extra bit to the message size.

The significance of the minimum value in the FP format of the log-LLR messages cannot be overstated, as it represents the smallest magnitude of LLR values. Consequently, the log-LLR values exhibit an uneven range around zero, with a negative offset. This imbalance necessitates an additional bit for the integer part of the FP format. To circumvent this issue, we introduce a constant positive offset to the messages, effectively eliminating the sign bit of the log-LLR and making room for the sign bit of the LLR value.

For example, suppose that the SPA requires $8$ bits of precision. The smallest non-zero value of $\gamma$ that this format can handle is $2^{-8}$, which corresponds to \mbox{$\ln(2^{-8}) \approx -5.5$} in log-LLR. To handle LLR values up to $10$, the log-LLR representation must cover the interval $[-5.5, 2.5]$. A possible way to represent this range is to use FP notation with one sign bit and three integer bits, which would need a total of $4$ bits. However, we can reduce the total bits to $3$ by adding a constant offset of $5.5$ to the log-LLR values, which removes the sign bit from the log-LLR format. This ensures that all log-LLR values are positive and lie in the range of $0$ to $8$. The modified decoding algorithm incorporates this offset operation.

Assuming that the added offset term is denoted as $b$, the updated decoding algorithm can be expressed as follows:
\begin{enumerate}
    \item Initialization: for each VN $i$ with $i \in \{ 1 \ldots N \} $ and $j \in \mathcal{N}(i)$, set $\dbtilde{L}^{[\mathrm{v}]}_{i \to j } = \tilde L_{\mathrm{ch},i} + b$ and \mbox{${\alpha}^{[\mathrm{v}]}_{i \to j } = \alpha_i = \sign(L_{\mathrm{ch},i})$}. 
    \item CN update: for each CN $j$ and \mbox{$i \in \mathcal{M}(j)$}, calculate
\begin{align}
    \dbtilde{L}^{[\mathrm{c}]}_{i \gets j } &= \dbtilde{L}^{[\mathrm{v}]}_{m \to j } + \hspace{-0.7cm} \sum_{\ell\in \mathcal{M}(j)\backslash \{i, m\}} \hspace{-0.5cm} \ln\left(  \tanh{\left(\frac{\exp{(\dbtilde{L}^{[\mathrm{v}]}_{\ell \to j } - b)}}{2}\right)} \right) \nonumber\\ 
    {\alpha}^{[\mathrm{c}]}_{i \gets j } &=  \prod_{k\in \mathcal{M}(j)\backslash \{i \}} {\alpha}^{[\mathrm{v}]}_{k \to j }, \nonumber
\end{align}
where $m$ is defined in (\ref{eqn:defn_m}).
    \item VN update: for each VN $i$  and \mbox{$j \in \mathcal{N}(i)$}, calculate
\begin{align}
    {L}^{[\mathrm{v}]}_{i \to j } &=  \alpha_i \exp(\dbtilde L_{\mathrm{ch},i} - b) + \hspace{-0.5cm} \sum_{k\in \mathcal{N}(i)\backslash \{j \}} \hspace{-0.35cm} {\alpha}^{[\mathrm{c}]}_{i \gets k } \exp(\dbtilde{L}^{[\mathrm{c}]}_{i \gets k } - b ) \nonumber \\
     \dbtilde{L}^{[\mathrm{v}]}_{i \to j } &= \ln\left(\lvert {L}^{[\mathrm{v}]}_{i \to j } \lvert\right) + b \nonumber \\
       {\alpha}^{[\mathrm{v}]}_{i \to j }  &= \sign\left( {L}^{[\mathrm{v}]}_{i \to j }  \right) \nonumber
\end{align}
    \item A posteriori LLR: for each VN $i$ with $i \in \{ 1 \ldots N \} $, calculate
\begin{align}
    L^{[\mathrm{total}]}_{i } &= \alpha_i\exp(\dbtilde L_{\mathrm{ch},i} - b) + \hspace{-0.3cm}\sum_{k\in \mathcal{N}(i)} {\alpha}^{[\mathrm{c}]}_{i \gets k }  \exp(\dbtilde{L}^{[\mathrm{c}]}_{i \gets k } - b) \nonumber \\
    \hat{x}_i &= \frac{1- \sign{(L^{[\mathrm{total}]}_{i })}}{2} \nonumber
\end{align}
    \item Stopping Criteria:  if $\bold{H\hat{x}}^T=\bold{0}$, with $\bold{\hat{x}} = (\hat{x}_1, \ldots, \hat{x}_N )$ or the number of decoding iterations equals the limit, stop. Otherwise, go to Step 2.
\end{enumerate}

We start by computing the logarithm of the LLR values of the channel and adding the offset term $b$.  We can use the log-sum/difference-exp function for the VN update, as described in Section \ref{sec:log_log_SPA}. The CN update requires evaluating \mbox{$\ln(\tanh(\exp(x-\ln(2)-b)))$}, which can be simplified as $g(x-b)$ in (\ref{eqn:piecewise_approx}). We do not need to subtract or add $b$ in CN or VN update, because the output log-LLR values are offset by $b$ when input values are offset.

\section{Results}
\label{sec:results}

\subsection{Performance}

\begin{figure}[t]
\centering
\tikzsetnextfilename{output-figure4}%
\begin{tikzpicture}[every pin/.style={fill=white},scale=0.8]
  \begin{axis}[
    grid=major,
    xlabel={ \large \text{$E_\text{s}/N_0$} (dB)},
    ylabel={\large Frame Error Rate (FER)},
    ymode=log,
    legend pos=south west,
    grid=major,
    ymax=1,
    ymin=0.01,
    legend style={align=left},
    legend cell align=left, 
    xmin=-21.55,
    xmax=-21.1,
    axis line style =very thick,
  ]
\addplot[KITblue,mark=o, line width=1.5] table [x=SNR,y=FER] {data/R_0p01_SPA_double_precision.txt};
\addlegendentry{ \small SPA - floating point} %
\addplot[KITred,mark=square, line width=1.5] table [x=SNR,y=FER] {data/FP_SPA_3_4_8_R_0p01.txt};
\addlegendentry{ \small SPA - FP(1,3,8)} %
\addplot[KITgreen,mark=square, line width=1.5] table [x=SNR,y=FER] {data/FP_log_3_4_6_5_R_0p01.txt};
\addlegendentry{ \small log-LLR - FP(1,3,6)} %
\end{axis}
\end{tikzpicture}
\captionsetup{font=small}
\caption{Performance comparison of floating point sum-product algorithm (SPA), fixed point (FP) SPA and FP log-LLR decoding for the TBP-LDPC codes of rates $R=0.01$ with $N=998400$. FP(1,$x$,$y$) represents one sign bit, $x$ integer bits, $y$ fraction bits for FP
 representation.}
\label{fig:log_LLR_R_0p01}
\end{figure}
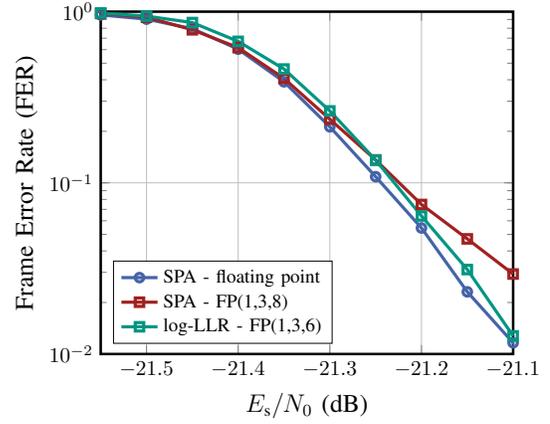
We demonstrate the resource reduction achievable with LDPC decoding using log-LLR messages by decoding the TBP-LDPC codes of rates $R=0.01$ and $R=0.1$ with the protographs in \figurename~\ref{fig:protographs}. We compare the performance with a fixed-point SPA decoder that processes LLRs, following the algorithm in Section \ref{sec:SPA}. In the log-log domain SPA decoder, we optimize the offset value $b$ in the fixed-point representation. We select $b=5$ after extensive simulations, which balances the representation of small and large LLR values within the fixed integer bit width.

\figurename~\ref{fig:log_LLR_R_0p01} shows the Monte-Carlo simulation results for the TBP-LDPC code of rate $R=0.01$. The log-LLR representation with $6$ fractional bits achieves a performance close to the floating-point SPA. The FP SPA with $8$ fractional bits suffers from saturation effects at low FER values, while the log-log domain SPA avoids them. The log-log domain SPA outperforms the FP SPA decoder in the ultra-low rate regime, even with a $25\%$ reduction in fractional bit width.

\begin{figure}[t]
\centering
\tikzsetnextfilename{output-figure5}%
\begin{tikzpicture}[every pin/.style={fill=white},scale=0.8]%
\begin{axis}[
    xlabel={ \large \text{$E_\text{s}/N_0$} (dB)},
    ylabel={\large Frame Error Rate (FER)},
    ymode=log,
    legend pos=south west,
    grid=major,
    ymax=1,
    ymin=0.0001,
    legend style={align=left},
    xmin=-11.3,
    xmax=-10.7,
    legend cell align=left, 
    axis line style =very thick,
]
\addplot[KITblue,mark=o, line width=1.5] table [x=SNR,y=FER] {data/R_0p1_SPA_double_precision.txt};
\addlegendentry{ \small SPA - floating point} %
\addplot[KITred,mark=square, line width=1.2] table [x=SNR,y=FER] {data/FP_SPA_3_4_7_R_0p1.txt};
\addlegendentry{ \small SPA - FP(1,3,7)} %
\addplot[KITgreen,mark=square, line width=1.5] table [x=SNR,y=FER] {data/FP_log_3_4_4_5_R_0p1.txt};
\addlegendentry{ \small log-LLR - FP(1,3,4)} %

\end{axis}
\end{tikzpicture} 

\captionsetup{font=small}
\caption{Performance comparison of floating point sum-product algorithm (SPA), fixed point (FP) SPA and FP log-LLR decoding  for the TBP-LDPC code of rate $R=0.1$ with $N=128000$. FP(1,$x$,$y$) represents one sign bit, $x$ integer bits, $y$ fraction bits for FP
 representation.}
\label{fig:log_LLR_R_0p1}
\end{figure}
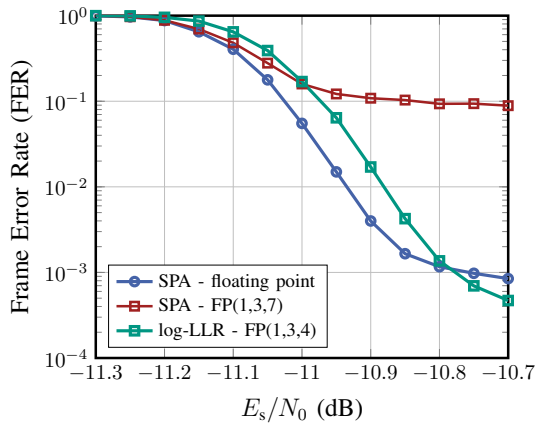

\figurename~\ref{fig:log_LLR_R_0p1} shows the simulation results for the TBP-LDPC code of rate $0.1$. The log-log domain SPA has a slight performance gap to the floating-point SPA due to the CN update approximation. However, this gap is only \SI{0.05}{dB} at an FER of $0.1$, which is acceptable as it only slightly affects the reconciliation efficiency $\beta$. The log-log domain SPA with $4$ fraction bits performs better than the FP SPA with $7$ fraction bits at an FER of $0.1$. The FP log-log domain SPA also has a lower error floor than the FP SPA, which saturates at an FER of $0.09$. The log-log domain SPA is a feasible option for CV-QKD and other low rate code applications, as it reduces the fractional bit width by more than $40\%$.

\subsection{Complexity}

We compare the complexity of different algorithms by examining their node update equations. In this analysis, we focus on pairwise message processing. The CN update equation for the SPA, as shown in (\ref{eqn:boxplus_section}), requires one minimum operation, two table lookups, two summations, and two subtractions. On the other hand, the VN update equation involves one summation for each LLR pair. For the log-log domain SPA, the CN update equation given by (\ref{eqn:CN_approx_piecewise}) involves one minimum operation, one summation, one subtraction, one multiplication, and up to three threshold comparisons. However, we do not consider the thresholds, scaling coefficients, and values to be subtracted in the decoding complexity, as they are constant and easily implemented in hardware, as discussed in Section~\ref{sec:log_log_SPA}. The VN update equation for the log-log SPA requires one maximum operation, one table lookup, one summation, and one subtraction for each LLR pair. We disregard the summations and subtractions, as they have lower complexity than the table lookups. Based on these considerations, we conclude that the log-log SPA and the SPA have similar complexities.

\section{Conclusion}
In this paper, we proposed a novel log-log domain SPA for decoding LDPC codes in CV-QKD systems. Our algorithm reduces the fractional bit width of the messages exchanged by the decoder by at least $25\%$, leading to lower memory consumption and simpler hardware implementation. We provided practical guidelines for fixed-point arithmetic and compared the performance and complexity of our algorithm with the conventional SPA. Our algorithm achieves decoding accuracy comparable to floating point SPA and superior to fixed point SPA.

 \bibliographystyle{IEEEtran}
 \bibliography{IEEEabrv,biblio}

\begin{thebibliography}{10}
\providecommand{\url}[1]{#1}
\csname url@samestyle\endcsname
\providecommand{\newblock}{\relax}
\providecommand{\bibinfo}[2]{#2}
\providecommand{\BIBentrySTDinterwordspacing}{\spaceskip=0pt\relax}
\providecommand{\BIBentryALTinterwordstretchfactor}{4}
\providecommand{\BIBentryALTinterwordspacing}{\spaceskip=\fontdimen2\font plus
\BIBentryALTinterwordstretchfactor\fontdimen3\font minus \fontdimen4\font\relax}
\providecommand{\BIBforeignlanguage}[2]{{%
\expandafter\ifx\csname l@#1\endcsname\relax
\typeout{** WARNING: IEEEtran.bst: No hyphenation pattern has been}%
\typeout{** loaded for the language `#1'. Using the pattern for}%
\typeout{** the default language instead.}%
\else
\language=\csname l@#1\endcsname
\fi
#2}}
\providecommand{\BIBdecl}{\relax}
\BIBdecl

\bibitem{Djordjevic_book}
I.~B. Djordjevic, \emph{Physical-layer Security and Quantum Key Distribution}, 1st~ed.\hskip 1em plus 0.5em minus 0.4em\relax Springer Nature, Sep. 2019.

\bibitem{208kmCV-QKD}
Y.~Zhang, Z.~Chen, S.~Pirandola, X.~Wang, C.~Zhou, B.~Chu, Y.~Zhao, B.~Xu, S.~Yu, and H.~Guo, ``Long-distance continuous-variable quantum key distribution over 202.81 km of fiber,'' \emph{Phys. Rev. Lett.}, vol. 125, p. 010502, Jun. 2020.

\bibitem{9562244}
K.~Gümüş and L.~Schmalen, ``Low rate protograph-based {LDPC} codes for continuous variable quantum key distribution,'' in \emph{Proc. International Symposium on Wireless Communication Systems (ISWCS)}, 2021.

\bibitem{Leverrier_CV_QKD_gaussian_mod}
P.~Jouguet, S.~Kunz-Jacques, and A.~Leverrier, ``Long-distance continuous-variable quantum key distribution with a {Gaussian} modulation,'' \emph{Phys. Rev. A}, vol.~84, p. 062317, Dec. 2011.

\bibitem{Mani_2021}
H.~Mani, T.~Gehring, P.~Grabenweger, B.~Ömer, C.~Pacher, and U.~L. Andersen, ``Multiedge-type low-density parity-check codes for continuous-variable quantum key distribution,'' \emph{Phys. Rev. A}, vol. 103, no.~6, Jun. 2021.

\bibitem{Milicevic_2018}
M.~Milicevic, C.~Feng, L.~M. Zhang, and P.~G. Gulak, ``Quasi-cyclic multi-edge {LDPC} codes for long-distance quantum cryptography,'' \emph{npj Quantum Information}, vol.~4, no.~1, Apr. 2018.

\bibitem{Cil24OFC}
E.~E. Cil and L.~Schmalen, ``Iteration-dependent scaled min-sum decoding for low-complexity key reconciliation in {CV-QKD},'' in \emph{Proc. Opt. Fiber Commun. Conf. (OFC)}, San Diego, CA, USA, Mar. 2024.

\bibitem{yangFPGABasedLDPCDecoder2021}
S.~S. Yang, J.~Q. Liu, Z.~G. Lu, Z.~L. Bai, X.~Y. Wang, and Y.~M. Li, ``An {FPGA}-based {LDPC} decoder with ultra-long codes for continuous-variable quantum key distribution,'' \emph{IEEE Access}, vol.~9, pp. 47\,687--47\,697, 2021.

\bibitem{zhouHighthroughputDecoderQuasicyclic2022}
\BIBentryALTinterwordspacing
C.~Zhou, Y.~Li, L.~Ma, J.~Yang, W.~Huang, H.~Wang, Y.~Luo, F.~C.~M. Lau, Y.~Li, and B.~Xu, ``High-throughput decoder of quasi-cyclic ldpc codes with limited precision for continuous-variable quantum key distribution systems,'' 2022. [Online]. Available: \url{https://arxiv.org/abs/2207.01860}
\BIBentrySTDinterwordspacing

\bibitem{silberhornContinuousVariableQuantum}
C.~Silberhorn, T.~C. Ralph, N.~L\"utkenhaus, and G.~Leuchs, ``Continuous variable quantum cryptography: beating the 3 db loss limit,'' \emph{Phys. Rev. Lett.}, vol.~89, p. 167901, Sep. 2002.

\bibitem{leverrierMultidimensionalReconciliationContinuousvariable2008}
A.~Leverrier, R.~Alléaume, J.~Boutros, G.~Zémor, and P.~Grangier, ``Multidimensional reconciliation for a continuous-variable quantum key distribution,'' \emph{Phys. Rev. Lett.}, vol.~77, no.~4, Apr. 2008.

\bibitem{Richardson:95821}
\BIBentryALTinterwordspacing
T.~Richardson and R.~Urbanke, ``Multi-edge type {LDPC} codes,'' 2004. [Online]. Available: \url{http://infoscience.epfl.ch/record/95821}
\BIBentrySTDinterwordspacing

\bibitem{maganaDifferentPerspectiveApproach2012}
M.~Magaña and P.~Poocharoen, ``\BIBforeignlanguage{en}{Different perspective and approach to implement adaptive normalised belief propagation-based decoding for low-density parity check codes},'' \emph{\BIBforeignlanguage{en}{IET Communications}}, vol.~6, no.~15, pp. 2314--2325, Oct. 2012.

\bibitem{Ryan2009}
W.~E. Ryan and S.~Lin, \emph{Channel Codes - Classical and Modern}.\hskip 1em plus 0.5em minus 0.4em\relax Cambridge University Press, 2009.

\end{thebibliography}

\end{document}